# NDE 4.0: The Fourth Revolution in Non-Destructive Evaluation: Digital Twin, Semantics, Interfaces, Networking, Feedback, New Markets and Integration into the Industrial Internet of Things

Johannes VRANA [1], DGZfP Subcommittee Interfaces [2]
[1] Vrana GmbH, Rimsting, Germany
[2] DGZfP, Berlin

Contact E-Mail: contact@vrana.net

**Abstract.** The industrial revolution is divided into three phases by historians: The invention of the steam engine (mechanization), electricity (mass production) and the microelectric revolution (automation). There was a similar development in non-destructive evaluation: tools such as lenses or stethoscopes allowed the human senses to be sharpened, the conversion of waves makes the invisible visible and thus offers a "look" into the components and finally automation, digitization and reconstruction. During the entire industrial development NDE was decisively responsible for the quality and thus for the success of the manufactured goods.

Industry is now talking about a fourth revolution: The informatization, digitization and networking of industrial production. As always, NDE will be critical to the success of this fourth revolution by providing the database needed for feedback in a networked production environment.

For NDE, this will lead to change. The test results must be made available to a networked production environment in such a way that they can be evaluated for feedback loops, the testability must be considered in the design and the reliability of the test statements will become increasingly important.

This publication presents first an orientation to NDE 4.0, including the development of Industry and NDE, a definition of its revolutions, a collection of several current-day challenges of NDE, and a discussion whether and how those can be solved with NDE 4.0.

Second this publication presents concepts on how NDE can be integrated into Industry 4.0 landscapes: The Reference Architecture Model Industry 4.0 (RAMI 4.0) shows the complete Industry 4.0 space and allows every Industry 4.0 standard and interface to be located. The Industry 4.0 Asset Administration Shell (AAS) implements the digital twin and is the interface between Industry 4.0 communication and the physical device. The Industrial Internet of Things (IIoT) ensures that different Industry 4.0 protocols can be combined using gateways. OPC UA is the communication protocol that is currently established as the standard and DICONDE is a communication protocol and data format for test data and metadata. Semantic interoperability is the basis to guarantee that all components can understand the information from all other components, and the International Data Spaces Association (IDSA) ensures data sovereignty, enables data markets and connects the world.

**Notes**

This paper presents the early stages on the road to Industry 4.0 and NDE 4.0.

The beginning of this paper shows in the quite extensive introduction the development of Industry and NDE, defines its revolutions, and collects several current-day challenges of NDE and discusses whether and how those can be solved with NDE 4.0. Therefore, this first part can be seen as an orientation to NDE 4.0 with a specific purpose to bring awareness and familiarity with the subject.

The second part goes into the details of one of NDE 4.0's largest business cases: NDE for Industry 4.0. It contains details from the Industry 4.0 / IT world which might be unfamiliar and therefore harder to understand for some NDE engineers and inspectors. However, with the NDE 4.0 development it will get increasingly important to understand the basics of this Industry 4.0 / IT world.

**Introduction**

The term Industry 4.0 was created in 2011 and has led to an almost unmanageable number of activities over the past 8 years. Thousands of people are working to make the dream of a networked industry come true thanks to open interfaces.

As an integral part of industrial production and operation, NDE (non-destructive evaluation) provides the quality assurance means required by industry. With the foundation of the DGZfP (German Society for NDE) committee "ZfP 4.0" in 2017, of the ASNT (American Society for NDT) committee "NDE 4.0" in 2019, and of the ICNDT (International Committee for NDT) Specialist International Group "NDE 4.0" in 2019 the NDE industry reacted to developments in connection with Industry 4.0. In addition, the DGZfP Subcommittee Interfaces and Documentation for NDE 4.0 faces the challenge of defining the interfaces between NDE and industry in such a way that customers can process and interpret NDE results directly in their world (Vrana 2019, Vrana 2019, Vrana 2019). *The NDE sector will not succeed in giving the industry new interfaces. It is more reasonable to use the Industry 4.0 interface developments and to participate in the design in order to shape them for the NDE requirements.*

*The Industrial Revolutions*

The terms Industry 4.0, Industrial Internet of Things (IIoT) and digital factory are now ubiquitous, but what do they mean? Industry 4.0 is the fourth industrial revolution, the IIoT one of the technologies that enables the connections necessary for this fourth revolution, and the digital or smart factory the goal. The term "4.0" refers to the version numbering common for software. The following is a brief overview of the four industrial revolutions (see also table 1).

The industrial revolution began in England in the second half of the 18th century and brought a change from handcrafted forms of production to the mechanization of production with steam engines or regenerative energy sources such as water.

The second industrial revolution was marked by the economic use of new chemical and physical knowledge and the beginning of new industries such as the chemical and pharmaceutical industries, electrical engineering and mechanical engineering. It began at the end of the 19th century in Germany and led to the introduction of the assembly line (1913 at Ford) and to new forms of industrial organization.

**Table 1:** The four industrial revolutions

|  | **Handcraft** | **Industry 1.0** | **Industry 2.0** | **Industry 3.0** | **Industry 4.0** |
|---|---|---|---|---|---|
| Revolutionary innovations |  | Simple mechanization | New industries, mass production | Computer & automation | Networking, data market |
| Key enablers | Fire, tools | Steam engine, renewable energies | Chemical and physical findings, production line | Digital technology, robots, drones | Informatization, digitalization, networks, interfaces, digital communication, artificial intelligence, machine learning, 5G, quantum technologies |
| Technological basis | Muscle power | Coal, iron | Electricity | Microelectronics | Software, computer science |
| Leading country |  | England | Germany | USA | ? |

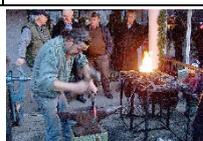
© Johann Jaritz

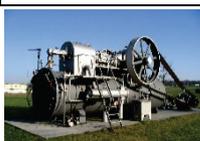
© Wassily Frese

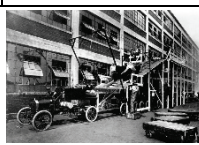

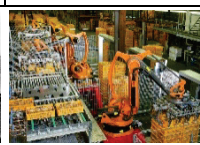

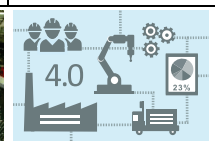
© Franziska Vrana

At the end of the 20th century, the development of microelectronics, digital technology and computers ushered in the third industrial revolution, which allowed automated control of industrial production and revolutionized data processing in offices (computers, laptops) as well as in private environments (computers, mobile phones, game consoles).

All these developments were enabled by the emerging technologies of the particular period, were implemented to simplify industrial production and allowed new and cheaper products. For example, the textile industry was started with the first revolution which allowed everybody to afford clothing. However, multiple professions became unnecessary and working conditions challenging. Which, in the long run, resulted in the creation of trade unions, better and safer working conditions, more jobs, shorter worktime, longer life expectation, and a higher living standard for everybody. The second and third revolution helped to build further industries, made more products affordable (or enabled them like a computer), made more professions and certain product categories unnecessary but in the long run they improved the working and living conditions, created jobs and resulted in higher living standard up to the point that a 40-hour work week and an expected lifetime of 80 years is nowadays considered normal.

Nowadays a couple of new developments, like informatization, digitization, digitalization, networking, and semantic interoperability are changing/simplifying everybody's life and enable new products, for example web mapping tools (like Goggle maps), self-driving vacuum cleaners or cars, intelligent virtual assistants (like Amazon's Alexa), cryptocurrencies (like bitcoin), or ridesharing companies (like Uber). All those new products can be seen as the outcome of the ongoing fourth industrial revolution.

- Informatization is the process by which information technologies, such as the World Wide Web and other communication technologies, have transformed economic and social relations to such an extent that cultural and economic barriers are minimized (Kluver 2000).

- Digitization is the transition from analog to digital.

- Digitalization is the process of using digitized information to simplify specific operations.

- Networking uses digital telecommunications networks for sharing resources between nodes, which are machines / assets / computers that use common wire-based or wireless telecommunications technologies. To allow straightforward communication between the nodes it is best to use standard open interfaces. Those interfaces will be discussed later in this document.

- Semantic interoperability allows nodes to understand the received data and makes it machine readable.

The enablers for those developments, the emerging technologies of the fourth industrial revolution, are for example:

- new communication channels, such as 5G
- new computer technologies for evaluation, like GPGPUs (General Purpose Computation on Graphics Processing Unit), single-board computers, special hardware for AI (Artificial Intelligence) calculations and quantum computers
- and new ways to protect data from manipulation, such as quantum cryptography and blockchains.

Taking the example of the self-driving car. The car uses the data from multiple sensors to determine for example its position and its distance to other cars. Therefore, networks between all the sensors and the central computer must be established. By choosing open standardized interfaces the car manufacturer only needs to implement the standard interface once and afterwards he can use all sensors. Moreover, due to the semantic interoperability the car knows that the sensor is measuring a distance and that it is located at the front of the car. In addition, the car gets maps and traffic conditions from web mapping tools and gets information from the cars around itself from all kinds of manufacturers. All this combined information finally allows the car to choose and ride the way to a certain waypoint.

This shows the necessity of standardized interfaces for all kinds of systems, sensors, … and leads to the situation that sensor manufacturers insisting on using their own proprietary interfaces will soon not be used anymore – even that they might offer the best suited sensor in the market.

A similar development occurs in industrial manufacturing. Manufacturing shops are starting to collect the data from all kinds of manufacturing and handling machines, are installing sensors to monitor production, are connecting Enterprise Resource Planning and Manufacturing Execution Systems to simplify, enhance and secure industrial production, to streamline supply chains and to allow new, cheaper, and safer products. In addition, the wish that those cyberphysical make decisions independently is growing.

This results in the need for open standardized interfaces with semantic interoperability between all devices in the industry. To drive those developments the term Industry 4.0 was created in the year 2011. Within a very short time, especially in Germany, many projects and groups were created with the aim of standardizing the development, like the Platform Industry 4.0 and the International Data Spaces Association (IDSA). Without them, the fourth revolution cannot function. Similarly, the Industrial Internet Consortium (IIC) was established in the USA in 2014 working on the IIoT standards.

So, even that from a hardware standpoint the fourth industrial revolution uses the technical principles of the third revolution it leads to a completely new transparency of information through the informatization, digitalization and networking of all machines, equipment, sensors and people in production and operation. Industry 4.0 enables feedback and feedforward loops to be established in production, trends to be determined through data analysis and a better overview to be gained through visualization.

As already indicated, the first three industrial revolutions were declared by historians. The fourth, on the other hand, uses the term "4.0" to introduce it. For the reasons given above, however, it might be appropriate to speak already of a fourth revolution. However, only history will show whether it is worthy of the name.

*The Revolutions within Non-Destructive Evaluation*

Non-destructive testing and evaluation underwent a similar development compared to industry and can also be divided into four revolutions (see table 2). For the first industrial revolution, the basis was handcraft that had developed over the millennia. For NDE, the basis is perception. Through their senses, people have been able to "test" objects for thousands of years. They looked at components and joints, smelled, felt, tasted and knocked at them to learn something about their condition and interior.

The first revolution or the birth of non-destructive testing took place on the one hand through the introduction of tools that sharpened the human senses, and on the other hand through standardized tests. Procedures made the results of the tests comparable and tools such as lenses, colors or stethoscopes improved the detection capabilities. At the same time, industrialization also made it necessary to expand quality assurance measures.

The second revolution of NDE, like the second revolution of industry, is characterized by the use of physical and chemical knowledge and electricity. The transformation of electromagnetic or acoustic waves, which lie outside the range of human perception, into signals that can be interpreted by humans, resulted in a "look" into the components.

Parallel to industry, microelectronics, digital technology and computers made the third revolution in non-destructive testing possible. Digital inspection equipment, such as X-ray detectors, digital ultrasonic and eddy current equipment, and digital cameras have been developed, making it possible to automate inspection.

**Table 2:** The four revolutions of non-destructive testing after Vrana, Chancellor and Singh

|  | Perception | NDE 1.0 | NDE 2.0 | NDE 3.0 | NDE 4.0 |
|---|---|---|---|---|---|
| Revolutionary innovations |  | Procedures | "View" into components | Computer & automation | Networking, data market |
| Key enablers | Simple tools | Optical elements, soot, oil, chalk, colors, stethoscopes | Chemical and physical findings, e. g.: Ultrasonic & Electromagnetic waves (MT, ET, microwaves, terahertz, infrared, X-ray, gamma) | Digital technology, robots, drones, reconstruction | Informatization, digitalization, networks, interfaces, digital communication, artificial intelligence, machine learning, 5G, quantum technologies |
| Technological basis | Human Senses | Procedures | Electricity | Microelectronics | Software, computer science |

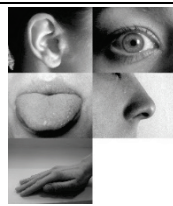 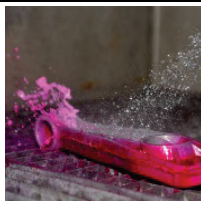 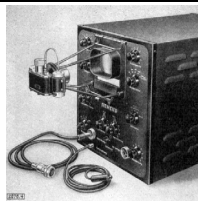 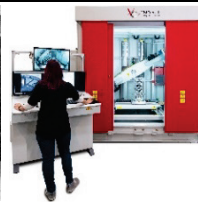 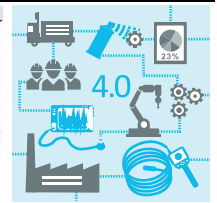

© Allan-Hermann Pool    © BMB    (Müller 1951)    © VisiConsult    © Franziska Vrana

The fourth revolution could become the greatest for non-destructive testing, turning the entire business upside down. First, the Industry 4.0 emerging technologies can be used to enhance NDE technologies and NDE data processing ("Industry 4.0 for NDE"). Second, a statistical analysis of NDE data provides insight into reliability, inspection performance, training status, consistency, and value of inspections ("Human Considerations"). Finally, NDE is the ideal data source for Industry 4.0 ("NDE for Industry 4.0") (Vrana 2020).

As with Industry 4.0, the aim is to create new information transparency through informatization and networking. *This will turn non-destructive evaluation from a niche product into one of the industry's most valuable sources of information.* This requires, however, like in the area of Industry 4.0, a standardization of the interfaces and the disclosure of data formats. Companies can now decide whether they want to follow the course of Blockbuster, Kodak, Quelle or Karstadt or rather follow Netflix and Google.

*Challenges of NDE*

To illustrate the benefits of NDE 4.0 a non-representative survey in social media was started (Facebook 2019, LinkedIn 2019). As the awareness of the benefits of Industry 4.0 and NDE 4.0 is not yet pronounced in practically any industry the question was regarding criticism on NDE and inspectors. From this it can be determined how NDE 4.0 can help to master these challenges. The responses listed below is the unfiltered and complete list of comments from the surveys in the social media and helped by showing a wide variety of challenges and identifying some necessary improvements. Some of the answers might present stereotypes, but even stereotypes contain a core truth. For a better understanding the responses have been grouped and editorial comments within the answers are indicated by brackets.

The following answers are related to criticism regarding education and morale in the NDE industry:

- ""NDE is not a skilled trade" is something I've heard over and over by some men in "skilled trades"."
- "Lack of
    - process knowledge
    - Surface preparation"
- "Operator dependent"
- "Reference is not up to the mark"
- "Risk outcomes for miss-calls in NDE are higher, making it more responsible and skill critical field whether its Aerospace, pipeline, or refinery work."
- "Each NDE methods own limitations for defect characterization make it harder for techs to master all methods to find all anomalies. UT expert may not confirm his finding by RT method since he is not expert in RT, make it more specific to individual with that skill. Which is hard for each tech to master all methods."
- "So many NDT inspectors who have not enough experience and little knowledge of welding making false calls"
- "Got offered something off the breakfast menu at McDonald's for me and my helper once on a turnaround. It was insulting because it was the ugliest weld I had ever seen on a 18in pipe and it was to 31.3 Severe Cycle too. It wouldn't have been as insulting had I been offered the dinner menu at least... Either way they had to cut it out cuz I don't do bribes […]"
- "I don't inspect chips"
- "Lack of ethics
    - in certification / qualification / training of technicians
    - in the application of test procedures
    - human factors are very important in risk-based management."

The second group of responses is related to the external perception or criticism of the benefits of non-destructive testing, or comments addressed to examiners:
- "Many times, other Engineers and project managers never include NDT Engineering in

planning because they believe they know everything there is to know about NDT. Many times, mindlessly prescribing methods that cannot detect the flaws or just throwing it in after planning with[out] even thinking. NDT Level IIIs and Engineers should always be included in design and planning phases. This will save money on the long run."

- "Why don't you inspect at a different location?"
- "Perform the spot test at a different location"
- "You mustn't look for indications in area you expect defects."
- "You can use another method, then the findings are acceptable."
- "The amount of welders who somehow think you have a magic pen for putting defects in radiographs is astounding. " That wasn't there when I welded it", says the welder with the X-ray vision!"
- "You don't need any inspection until something goes bang. Always chuckle when a welder tells you that they have never had a weld rejected. Two types of welders out there. There's those that accept that there's always a chance a weld will dip and there's those that tell a lot of lies."
- "We don't need NDT - you only test [and introduce] flaws into the material."
- "I got the "the other inspector never rejected anything, why are you rejecting so many pieces" guess something in the process changed is what I said."
- "NDT in civil engineering: "we don't need NDT, the safety factors in design will cover any flaws (and probabilities will cover any uncertainties)" and "If NDT becomes mandatory, our product will be too expensive for the market""
- "Its "no value added""
- "Production brake"
- "Turnover preventors"
- "Unnecessary cost factor"
- "You are like my mother in law, I don't need you... hate it when you are there... you create extra work for the rest of us and I end up paying a shitload of money"
- "NDT does not have any value at all. It only sorts out parts, that in reality are good. I don't want it and I would never ever do it, but my customer insists on it. I'd prefer spending the money into further improvement of my production!"

As an NDE sector, these points must be accepted as a point of view, evaluated and should be considered as opportunities for continual improvements in our field. The first group of answers is about training, morality, and reliability. These topics relate mainly to "Industry 4.0 for NDE" and "Human Considerations". "Industry 4.0 for NDE" could also be called emerging technologies for NDE and covers topics like the use of Artificial Intelligence (AI), Machine Learning (ML), Deep Learning (DL), big and smart data processing and visualization, cloud computing, Augmented / Virtual / Mixed Reality (AR / VR / MR), blockchains, 5G, quantum computers, enhanced robotics and drones, and revision-safe data formats and storage for a safer, cheaper, faster, and reliable inspection eco-system. "Human Considerations" covers topics like management and leadership 4.0, digital transformation and organization behavior, training and certification, standards and best practices, human factors, and human-computer interaction (HCI) (Vrana 2020).

The second group of responses shows that NDE is seen by many as an *unnecessary cost factor* and relate mainly to "NDE for Industry 4.0".

For a more detailed analysis of all of those answers refer to (Vrana 2020). This publication focusses on "NDE for Industry 4.0". "Industry 4.0 for NDE" and "Human Considerations" are not considered further in this publication.

*NDE 4.0 is the chance for NDE to free itself from this niche.* Until now, NDE methods have "only" been used to search for indications in order to meet standards that many customers think are unnecessary. But NDE can do more. NDE offers a view into the components and joints and is therefore an ideal database for digital engineering (Tuegel 2017), for better lifing calculations or fracture mechanical models (Vrana 2018), for the prediction of production problems, for the improvement of production, etc. This must be used. For this purpose, however, the *results of the NDE* must be made *available digitally* so that customers can evaluate the results. It therefore requires *standardized, semantic, manufacturer-independent interfaces and standardized open data formats*.

*This also requires a change of the thought processes of the inspectors.* Comments such as "I don't inspect chips" show that the concepts of Industry and NDE 4.0 need to be presented to the inspectors. In the context of Industry 4.0, all information is important. *Test results from areas that will later be machined also contain valuable information* that can be used, for example, to improve lifing models.

**Integration of NDE in the Product Development Process, in Production and in Operation and the Interfaces Specified Thereby**

As indicated above, NDE, as an integral part of the product development process, industrial production and industrial operation, provides the quality assurance means needed by industry.

During the product development process (see Figure 1), the specifications for production and inspection are created through the cooperation of experts from design, material sciences, production and NDE. These are inspected to optimize design and inspections. The value of NDE can already be seen here, as NDE offers a look into the prototypes and can therefore make a significant contribution to improving design and production. This requires interfaces for the statistical evaluation of the data (together with the process data from the inspections).

The data that can be obtained during the subsequent serial production and service give an even better picture of the components produced and their joints. This allows further improvements in design and production. In addition, they allow the next generation of products to be optimized (feedforward).

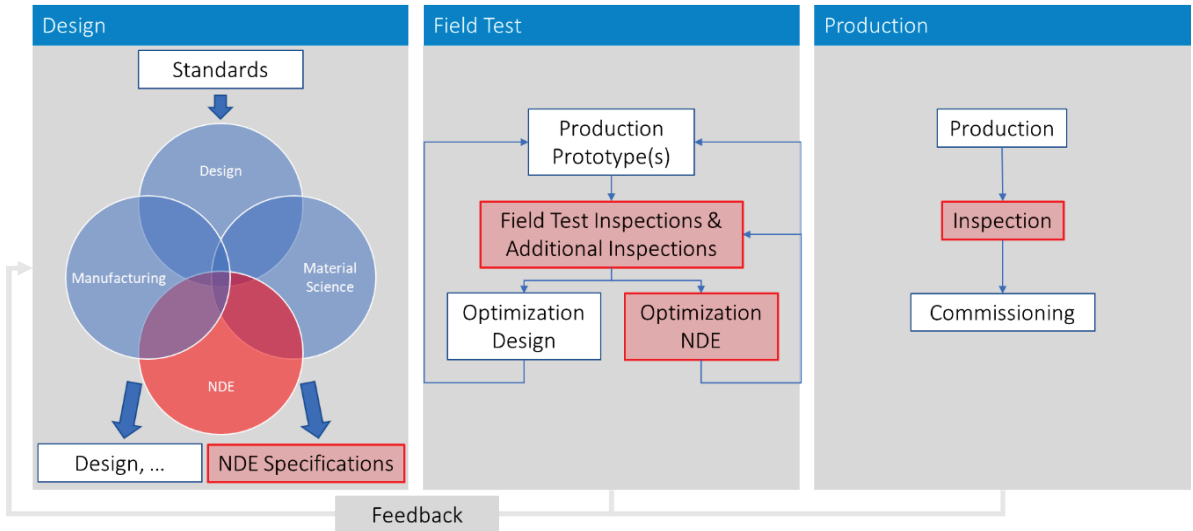

**Fig. 1** Typical product development process
(Figure 2 provides a more detailed description of the situation during inspection in serial production).

Figure 2 shows a closer look at the serial production and the inspections in the supply chain. Starting with material suppliers, who already carry out inspections on the raw material, through inspections at the component suppliers to the inspections at the OEMs, who assemble the final product. After all, the user is responsible for commissioning and service checks after certain operating times. All these inspections provide results that could be integrated into an Industry 4.0 world through appropriate interfaces and thus, as described above, could contribute to improving production and design.

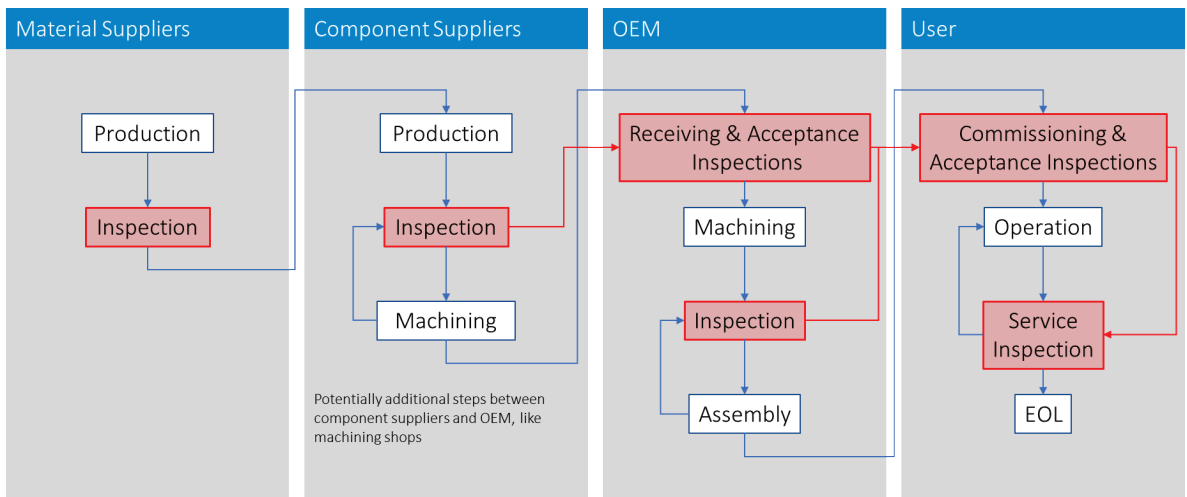

**Fig. 2** Typical supply chain with inspection steps in serial production.

Figure 3 shows the interfaces of each individual inspection step. The input interfaces marked in green supply the order data, provide the inspector with information on the component, serve to correctly set the devices, the inspection, the mechanics and the evaluation and to document the results in accordance with the specifications.

Digitalization of these input interfaces will help to support the inspector in his work, to avoid errors in the inspection, to optimize the inspection and to ensure a clear, revision-safe assignment of the results by digital machine identification of a component.

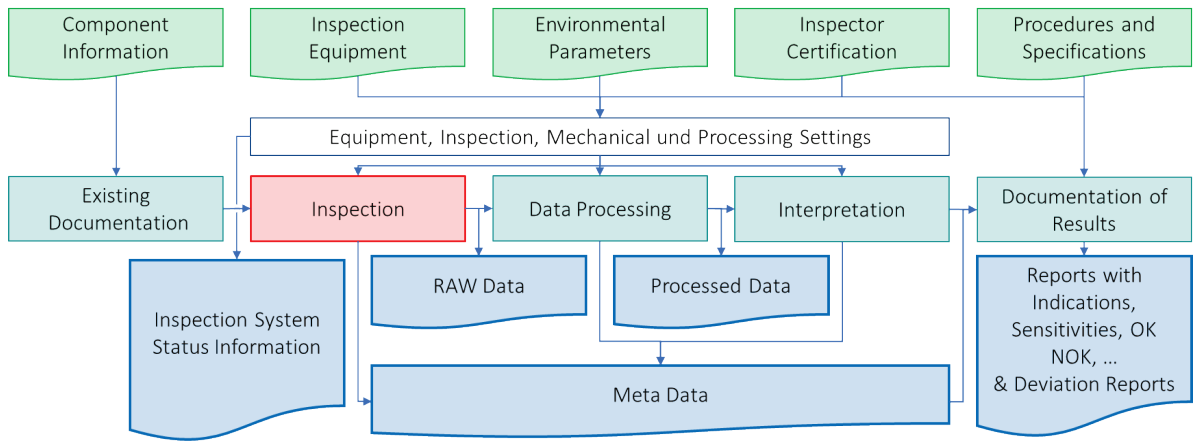

**Fig. 3** Typical sequence of an (automated) inspection in serial production
(can in principle be used for manual checks).

On the output side, the inspection system status information and the inspection results are generated. The inspection system status information could be used for maintenance and to improve the inspection system itself. The inspection results consist of the actual test data, the raw and processed data and the metadata (meaning the framework parameters of the inspection and evaluation), and finally the reported values. The reported values represent the key performance indicators (KPIs) of the inspection. For industry, interpreted data are the easiest to evaluate. Therefore, the reported values are currently the most relevant data of the inspection. Consideration should be given to whether the currently reported values are sufficient for NDE 4.0 purposes or whether the results to be reported should be extended for statistical purposes and thus for greater benefit to the customer.

**Automation Pyramid**

In a digitized industrial production environment ("Industry 3.0") the techniques and systems in process control are classified using the automation pyramid (see figure 4). The automation pyramid represents the different levels in industrial production. Each level has its own task in production, whereby there are fluid boundaries depending on the operational situation. This model helps to identify the potential systems / levels for Industry 4.0 and NDE 4.0 interaction (in particular regarding the beforementioned input and output parameters of an NDE inspection). However, validity of this model needs to be discussed in regard to Industry 4.0 and NDE 4.0.

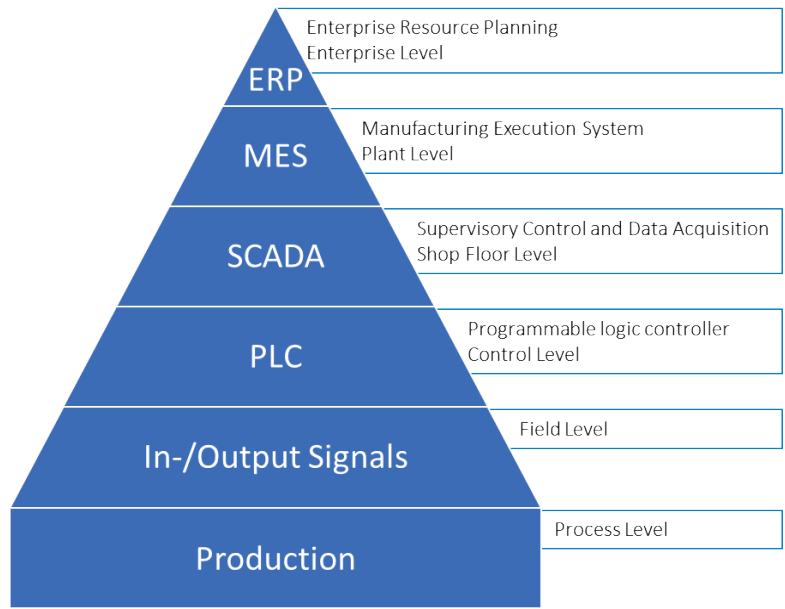

**Fig. 4** The Automation Pyramid.

Level 0 (process level) is the sensor and actuator level for simple and fast data collection. The field level is the interface to the production process using input and output signals. The control level uses systems like programmable logic controllers (PLC) for controlling the equipment. Supervisory control and data acquisition (SCADA) of all the equipment in a shop happens at shop floor level. Manufacturing execution systems (MES) are usually used for collecting all production data and production planning on the plant level. Finally, Enterprise Resource Planning (ERP) systems control operations planning and procurement for a company. Systems for Product Lifecycle Management (PLM) are usually not included in the automation pyramid (as the automation pyramid visualizes the automation during production and not during the lifecycle of a product) but are clearly connected to both the MES and the ERP systems.

The information flow for the planning of production comes from the ERP system and is broken down to the field/process level (meaning the communication starts at the top level of the pyramid and is communicated to the bottom layer). Once production is running the data is collected by the field/process level, is condensed in several steps (in the different levels), and finally the key-performance indicators (KPI) are stored in the ERP system (meaning the communication starts at the bottom levels of the pyramid and is communicated to the top level). For this information flow in both directions interfaces need to be implemented between the levels. Depending on the number of systems or devices in a level the number of interfaces to be implemented can be exhausting. This is why in a lot of production environments still analog (paper-based) or not-machine-readable digital (Email or PDF) solutions are used for certain interfaces between levels. However, such solutions require human action and are highly error-prone (like entering the 10-digit serial number of a certain component). This already shows the need for standard, machine readable interfaces.

In such an environment the main interaction system for NDE is the MES system, as this is the point where all the data from all the equipment is combined.

However, the idea of Internet 4.0 is not only to collect and analyze the data from all devices and systems (including PLM) but also that every device and system (including all NDE equipment) is able to communicate with each other device and system. All this independent of the level of the automation pyramid. Therefore, not only interfaces between two adjacent levels become necessary but interfaces between all devices and systems in all levels. This implementation effort for all the interfaces would prevent Industry 4.0. This is why standardized, open, and machine-readable interfaces become key for Industry 4.0 and this is why companies will have to shift from proprietary interfaces to standard interfaces if they want to survive the ongoing fourth industrial revolution. Looking onto the member lists of the ongoing standardization efforts shows that most of the big players (for example SAP, Microsoft, Siemens) are beginning to understand this. Unfortunately, a lot of small and medium companies are still ignoring this development.

**Digital Twin, Semantic Interoperability and Data Security**

Every asset, meaning every manufacturing device, sensor, product, software, person, operator, engineer, … can be described in the virtual world with information like shape, type, functionality, material composition, operational data, financial data, interfaces, etc.. All this information combined makes its virtual representation, the digital twin.

As discussed in the section above the data for the digital twin comes from all levels of the automation pyramid including the manufacturing execution system (MES) for all manufacturing related data, the enterprise resource planning (ERP) system for corporate data, and in addition the Product Lifecycle Management (PLM) for data from product development.

To create digital twins and for all Industry 4.0 communication it is important that the information is machine readable. It must be possible to interpret the meaning of the exchanged data unambiguously in the appropriate context. This is called semantic interoperability.

With the semantic information stored in the digital twin it will be possible to simulate the asset, to predict its behavior, to apply algorithms etc.. A digital twin can also include services to interact with the asset.

The user profiles and all the user activities maintained by social media platforms or the data stored about individuals by insurance companies, by companies, or by government can be seen as a part of a digital twin of a person. Already the data stored by one of those entities has quite some value. All the information combined in one digital twin would have an incredible value for certain entities but are a great threat for society as it leads to transparent humans. This shows the need for data security and sovereignty.

Data security is a means for protecting data (for example in files, emails, clouds, databases, or on servers) from unwanted actions of unauthorized users or from destructive forces. Therefore, data security is the basis for data-centric developments like the Industry 4.0 landscape discussed in this paper.

Data security is usually implemented by creating decentralized backups (to protect from destructive forces) and by using data encryption (to protect from unwanted actions). Data encryption is based on mathematical algorithms which encrypt and decrypt data using encryption keys. If the correct key is known encryption and decryption can be accomplished in a short time, but if the key is not known the decryption becomes very challenging for current-day computers (several months or years of calculation time) and the data is therefore secured from unwanted access. However, with computers becoming increasingly more powerful over time, encryption keys and algorithms need to become more challenging over time. And data encrypted with old algorithms or too short keys need to be re-encrypted after some time to keep it safe. The only measure ensuring data encryption over time is to use keys which have the same length as the data to be encrypted and which are purely random. One of the few methods to create such keys is quantum cryptography, which is still quite expensive in installation.

Where data security is the necessary basis, data sovereignty goes one step further protecting data. Data sovereignty guarantees the sovereignty of data for its creator or its owner. Data itself, if not artistic, is legally not protected by any copyright. Therefore, if a dataset is submitted to somebody else currently only individual contracts hinder the receiver from forwarding or selling the data (even if submitted using data encryption). Therefore, two measures have to be implemented to guarantee data sovereignty. #1 legal documents need to be prepared and #2 software and interfaces need to be implemented to restrict the use on receiving side to the rules of the submitting side.

In the industrial world data sovereignty is assured by measures like the ones discussed at the end of this publication. This enables the creation of reasonable digital twins, leads to added value, and creates new markets.

*Industry 4.0 Asset Administrative Shell (AAS)*

The Platform Industry 4.0 started the development of the Industry 4.0 Asset Administration Shell (AAS) (Plattform Industrie 4.0 2016, Plattform Industrie 4.0 2018) in 2015. The AAS is the virtual representation of each asset, its digital twin. An asset can be a device, but also a component, a plant, an entire factory, a software, or even a person / operator / inspector.

Each AAS (see Figure 5) consists of a manifest and a component manager. The manifest is a table of contents that provides all information about the asset in the header. In the body the manifest references all data stored by the asset and all functions that can be performed by the asset. The manifest is defined in XML or JSON (Plattform Industrie 4.0 2018). The component manager contains the actual implementations and realizes the interaction, functionality and high-performance data queries.

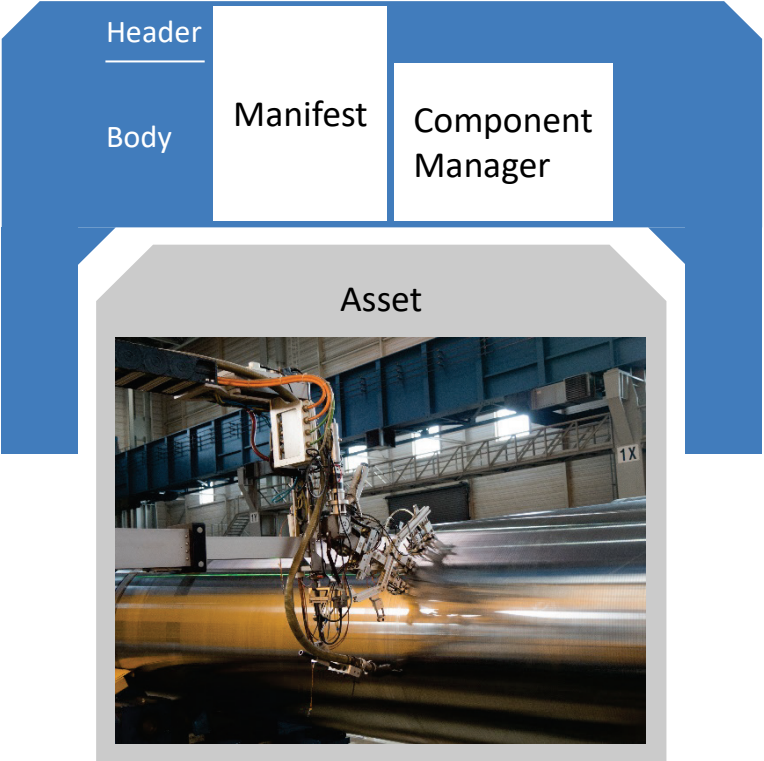

**Fig. 5.** Industry 4.0 Asset Administration Shell for an ultrasonic testing system.

Each AAS and each individual asset must have a globally unique identifier (ID), which is stored in the header. The ID of the AAS is the ID of the type – meaning whether it is a drill or a conveyor belt. The ID of the asset is the ID of the instance – meaning whether it is, for example, drill #1, #2, or #25.

AASs may be nested within each other. The AAS for a production line can reference the AAS of the various processing machines, inspection machines, etc.. And the AAS for an inspection system can, for example, contain the AAS for the mechanical drives, for the sensors and for the actual test system.

People, i.e. operators or inspectors, are also represented by an Asset Administration Shell. For example, there may be an AAS for a level 3 ultrasonic inspector specializing in the inspection of castings. This inspector receives his task via a tablet or an augmented reality platform and the results are stored digitally by the inspector. This shows that *Industry 4.0 is NOT striving for the deserted factory*. For Industry 4.0, networking is crucial and the results must be available digitally. It does not require automation. For some work steps, especially repetitive tasks, it makes more sense to use automated solutions. In other work steps the human being is more effective.

**Interfaces**

The introduction showed the need for standardized, vendor-independent interfaces and the AAS provides a standardized virtual representation of each asset describing the functionality and interfaces offered by the asset. But what are the interfaces in this context? Is it the question regarding the physical interface? The question regarding USB, WIFI or 5G? The question regarding TCP/IP, http, XML, or OPC UA? Before further discussion, the term interface must be specified in more detail.

*OSI (Open Systems Interconnection) Model*

The OSI model, see Figure 6, gives an overview of the different abstraction layers of digital interfaces and helps to select the interfaces that are decisive for NDE 4.0. The lowest level represents the physical connection, i.e. the cable or the radio connection. The first OSI layer, the transmission of the individual bits, runs via this connection. The information to be transmitted is combined with transmitter and receiver addresses and other information in the data link layer to form frames. Information packets are "tied" in the network layer and combined into segments in the transport layer.

The layers above are the so-called host layers. The session layer is responsible for process communication. The presentation layer is responsible for converting the data from a system-independent to a system-dependent format and thus enables syntactically correct data exchange between different systems. Tasks such as encryption and compression also fall into this layer. Finally, the application layer provides functions for applications, for example with application programming interfaces (API).

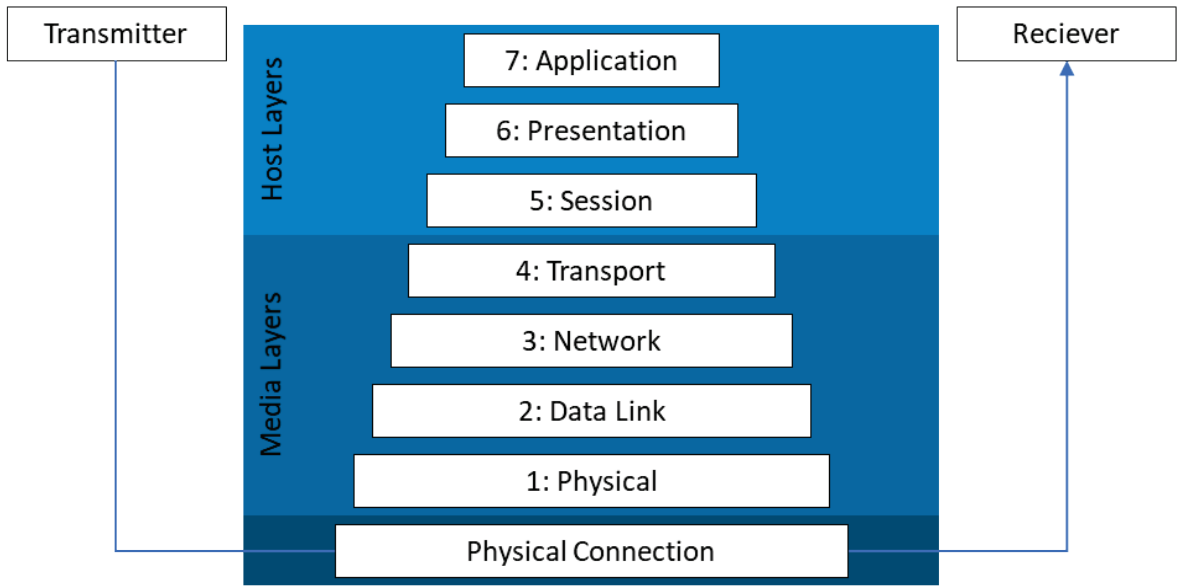

**Fig. 6** The OSI layers - a model for visualizing the degree of abstraction of interfaces

The application layer is the communication layer that is decisive for Industry and NDE 4.0. However, semantic interoperability (not to be confused with syntactic) needs be added on top for an appropriate Industry 4.0 communication. The physical connection (USB, WLAN, 5G, ...) is irrelevant.

An example of an application layer protocol is HL7 (Health Level 7). HL7 is the protocol used in healthcare to ensure interoperability between different information systems. HL7 (besides DICOM - see below) should therefore one of the interfaces for Medicine 4.0 and the communication can run over various physical connections. Other protocols such as OPC UA, Data Distribution Service (DDS) or oneM2M are gaining ground in the industrial world.

*Industrial Internet of Things*

The Industrial Internet Consortium (IIC) defines the Industrial Internet of Things (IIoT) in its specifications. In Volume G5 (Industrial Internet Consortium 2018) Internet 4.0 interfaces are discussed. Those discussions are based on the Industrial Internet Connectivity Stack Model, which is similar to the OSI model, however compared to the OSI model it combines the three host layers to one framework layer. Based on this model it compares the interface protocols OPC UA, DDS and oneM2M with Web Services (see figure 7). Every interface protocol is considered a Connectivity Core Standard and the need for Core Gateways between the Connectivity Core Standards is emphasized. This brings the benefit that every connectivity standard can be used, and the information combined using the gateways between the standards.

DDS is managed by Object Management Group (OMG) and focusses on low-latency, low-jitter peer-to-peer communication with a high Quality of Service (QoS). It is data-centric and does not implement semantic interoperability. There are plans to integrate DDS into OPC UA to integrate OPC UA Pub/Sub.

OneM2M is a connectivity standard mainly for mobile applications with intermittent connections and low demands regarding latency and jitter. Semantic interoperability is planned.

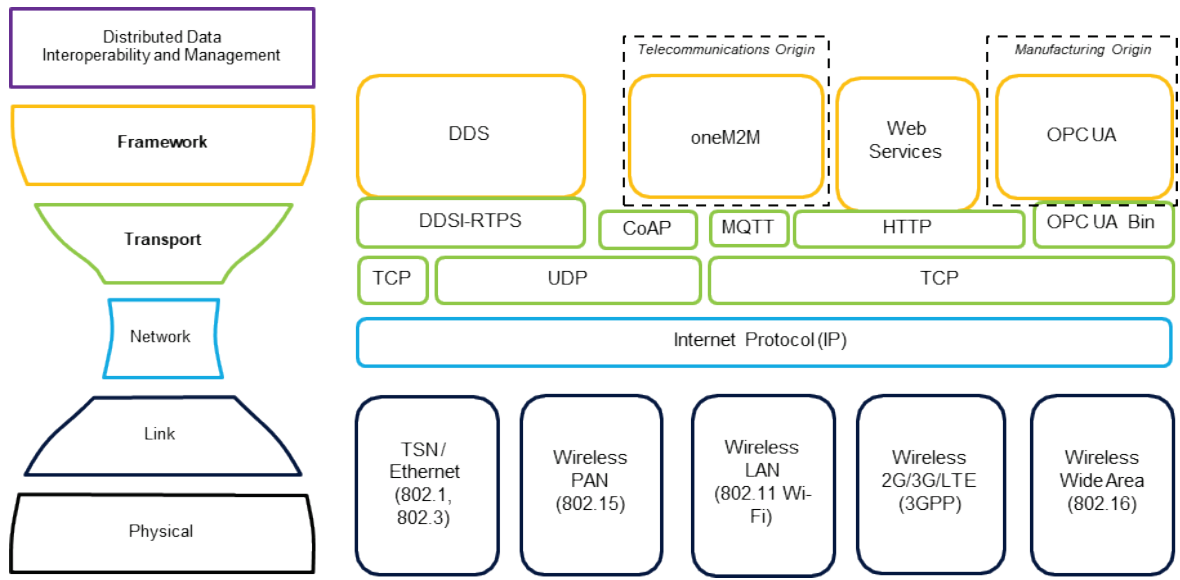

**Fig. 7** IIoT connectivity standards.
OPC UA has a manufacturing origin and oneM2M a telecommunication origin but both are now used for multiple industries, like DDS or WebServices. Transports that are specific to a connectivity standard are shown without any spacing between the framework and the transport layer boxes. (Industrial Internet Consortium 2018)

WebServices use the Hypertext Transfer protocol (HTTP) known from the internet. It is primarily for human user interaction interfaces. Semantic interoperability can be reached using the Web Ontology Language (OWL).

OPC UA, discussed in detail below, is mainly used in the manufacturing industry. In contrast to DDS it is object oriented and provides semantic interoperability.

For NDE applications OneM2M could be of benefit for mobile devices. WebServices are ideal for human-computer interaction and could be used for operator interfaces to store and read information regarding the component to be inspected. Low-latency and low-jitter communication is not necessary for typical NDE equipment; therefore, DDS will not be considered further. OPC UA, being the standard protocol for manufacturing and due to the included semantic interoperability, seems like the ideal interface for NDE 4.0.

*OPC UA*

The high-level communication protocol / framework that is currently established in the manufacturing Industry 4.0 world is OPC UA (OPC Foundation 2018, IEC 62541). OPC UA has its origin in the Component Object Model (COM) and the Object Linking and Embedding (OLE) protocol. OLE was developed by Microsoft to enable users to link or embed objects created with other programs into programs and is used extensively within Microsoft Office. COM is a technique developed by Microsoft for interprocess communication under Windows (introduced in 1992 with Windows 3.1). This standardized COM interface allows any program to communicate with each other without having to define an interface separately. With the Distributed Component Object Model (DCOM) the possibility was created that COM can also communicate via computer networks.

Based on these interfaces, a standardized software interface, OLE for Process Control (OPC), was created in 1996, which enabled *operating system independent* data exchange (i.e. also with systems WITHOUT Windows) in automation technology between applications from different manufacturers.

Shortly after the publication of the first OPC specification, the OPC Foundation was founded, which is responsible for the further development of this standard. The first version of the OPC Unified Architecture (OPC UA) was finally released in 2006. OPC UA differs from OPC in its ability not only to transport machine data, but also to describe it *semantically in a* machine-readable way. At the same time, the abbreviation OPC was redefined as Open Platform Communications.

OPC UA uses either TCP/IP for the binary protocol (OSI layer 4) or SOAP for web services (OSI layer 7) (see also Figure 6 and 7). Both Client-Server and Pub-Sub architectures are supported by the OPC UA communication framework. Based on this, OPC UA implements a security layer with authentication and authorization, encryption and data integrity through signing. APIs (Application Programming Interfaces) are offered to easily implement OPC UA in programs. In the .net framework OPC UA is even an integrated component. This means that the users do not have to worry about how the information is transmitted. This is done completely in the OPC UA framework (referred to as Infrastructure in Figure 8). The only thing that matters is what information is transmitted.

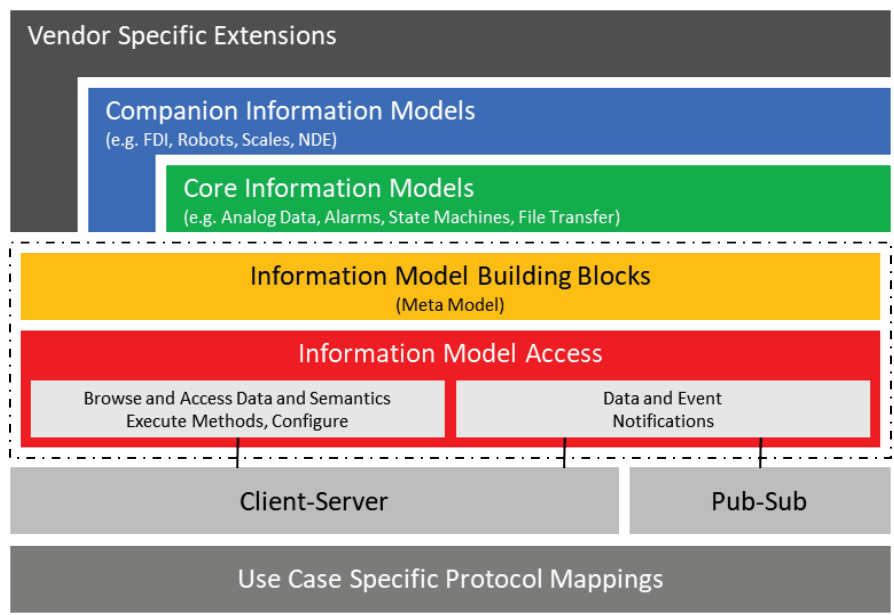

**Fig. 8** OPC UA architecture (© Vrana GmbH, based on OPC Foundation 2019)**.**

As Figure 8 shows, the OPC information model already defines some basic core information models in which models are defined that are required in many applications. In addition, companion specifications exist for product classes such as field devices (FDI), robots or scales. These companion specifications provide semantic interoperability and are therefore the basis for Industry 4.0, the basis for smooth I4.0 interfaces and communication and result in any OPC UA-enabled device being able to interpret data from others. In addition, there may also be manufacturer-specific specifications for the exchange of data between the devices of one manufacturer.

OPC UA Pub/Sub integrates DDS into OPC UA to enable One-to-Many and Many-to-Many communications. Moreover, OPC UA TSN (Time Sensitive Network) will make it possible to transfer data in real time and to extend OPC UA to the field level. The OPC UA specifications are also currently being converted into national Chinese and Korean standards.

Moreover, it is planned to start the development of an NDE companion specification for OPC UA in a joint project between DGZfP, VDMA and OPC Foundation.

OPC UA is, like HL7 in healthcare, the standard for an interface to the manufacturing Industry 4.0 world. In the same way as in medical diagnostics, large amounts of data are in some cases generated with NDE (in OPC UA larger files are split into smaller packages – e.g. the OPC UA C++ Toolkit has a maximum size of 16 MB). Computed tomography (CT), automated ultrasonic testing and eddy current testing can easily result in several GB per day that need to be archived long term. In the healthcare sector those large data files resulted in the development of DICOM (Digital Imaging and Communications in Medicine) alongside HL7.

*DICOM*

DICOM is an open standard with semantic interoperability for the storage and communication of documents, image, video and signal data and the associated metadata as well as for order and status communication with the corresponding devices. This will enable interoperability between systems from different vendors, as Industry 4.0 is striving for.

In health management, this leads to the necessity of interfaces between HL7 and DICOM (see Figure 9). This interface is usually found in the PACS (Picture Archiving and Communication System) server. In the process, patient and job data are translated from HL7 to DICOM for communication to the imaging devices. Information about the order status, about provided services (e.g. "X-ray image of the lung ") as well as written findings and storage locations of the associated images are communicated back. The returned data, texts and references would usually be referred to in industry as KPIs (Key Performance Indicators).

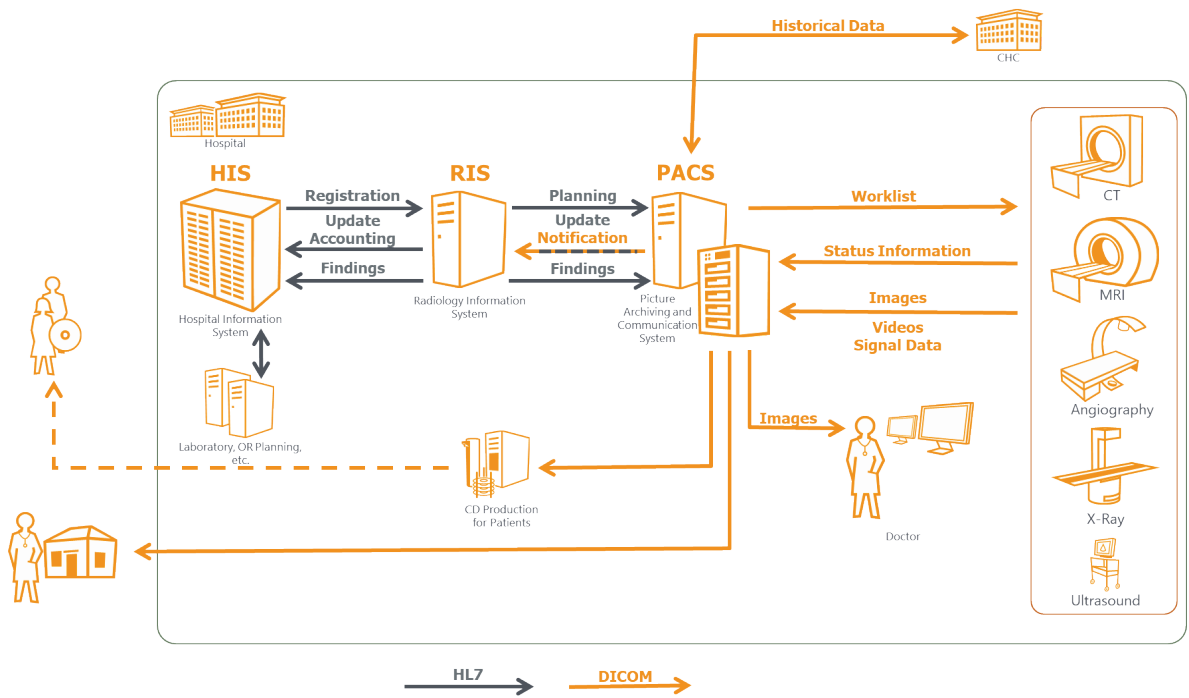

**Fig. 9** Interaction between HL7 and DICOM (©VISUS Industry IT GmbH, Germany).

The central system for the "process logic" in hospitals is the HIS (Hospital Information System; comparable to an ERP system in industry), which communicates with all other systems via HL7. All image, video and signal data are stored in DICOM format in

PACS, which is designed to handle large amounts of data and is the central system for archiving and communicating the data.

*Digital Workflow in NDE with OPC UA and DICONDE*

For the NDE world, this system can be transferred from HL7 and DICOM as follows (see Figure 10): The Industry 4.0 world consists of ERP (Enterprise Resource Planning) or MES (Manufacturing Execution System) servers for production planning or as a production control system and assets supply data via OPC UA. A transmission of order data for inspections as well as a return transmission of notifications and inspection results (KPIs for storage in the MES) can be mapped via OPC UA. An integration of maintenance and calibration data of NDE equipment via OPC UA is also conceivable.

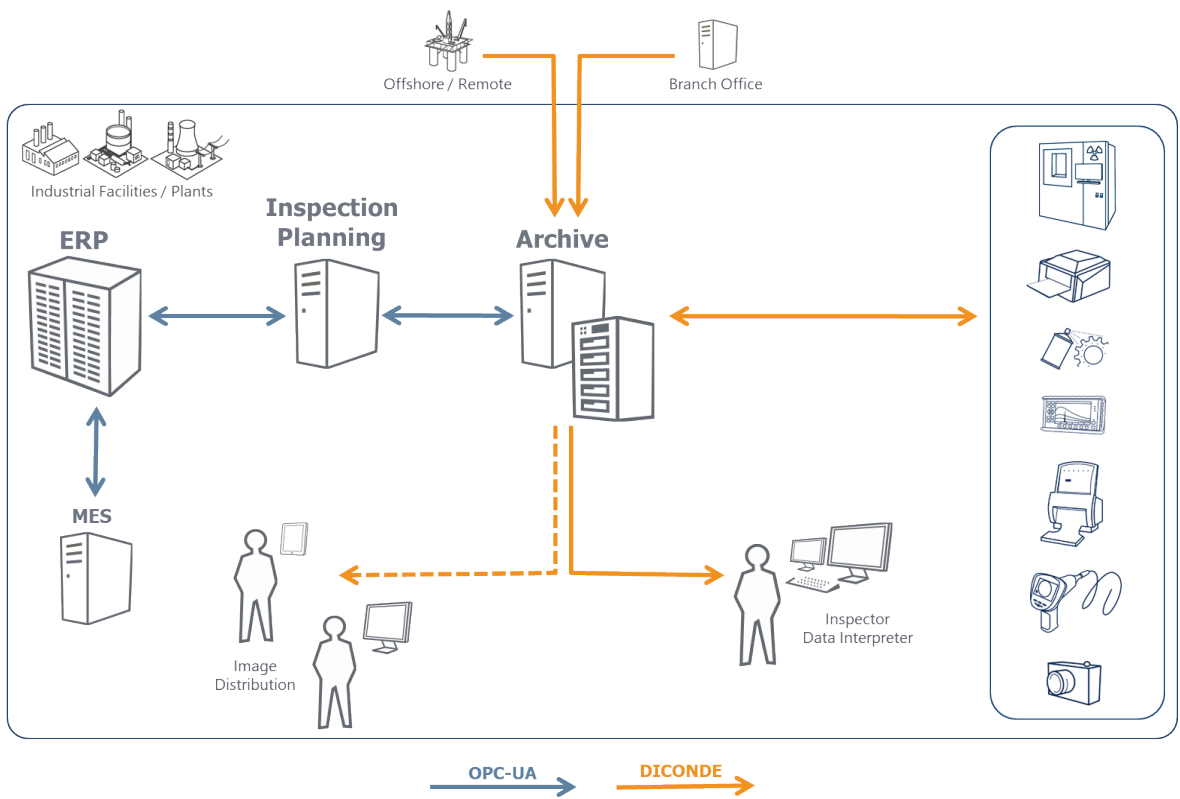

**Fig. 10** Possible interaction between OPC UA and DICONDE (©VISUS Industry IT GmbH, Germany).

With a few exceptions, however, the raw data generated during tests are too large to be communicated via OPC UA. Like HIS in a hospital, ERP and MES are not designed for the administration, communication and archiving of large amounts of image, video or signal data, such as is generated in radiography, computed tomography, automated ultrasound and eddy current testing or SAFT/TFM. Therefore, it makes sense to store the raw data outside the OPC UA world in a revision-proof way. The DICONDE standard offers itself as protocol and data format offering semantic interoperability. DICONDE is based on DICOM and has been adapted by ASTM to the requirements of the various NDE inspection methods (ASTM E2339 2015, ASTM E2663 2018, ASTM E2699 2018, ASTM E2738 2018, ASTM E2767 2018, ASTM E2934 2018). In radiography the DICONDE standard fits very well to the requirements of the users. There are already many manufacturers who store their data in the DICONDE format and have implemented the DICONDE communication interfaces, for example for the digital query of inspection orders, whose IDs are then automatically stored in the metadata of the DICONDE files and thus ensure structural integrity between NDE raw data and ERP/MES. DICONDE is also currently established as the standard in the field of computer tomography. Similar to healthcare, an entity that "translates" order data and reported values between OPC UA and DICONDE makes sense.

In ultrasonic and eddy current testing, however, the medical requirements are further apart from the requirements of NDE. Although the DICONDE standard strives to define suitable data formats (ASTM E2339 2015, ASTM E2663 2018, ASTM E2699 2018, ASTM E2738 2018, ASTM E2767 2018, ASTM E2934 2018), these are currently not supported by device manufacturers. It is necessary to clarify at which points the manufacturers still see a need for action.

Contrawise, DICONDE can be easily implemented for the connection of visual inspections, e.g. photos in the field of dye penetrant and magnetic particle inspection and videos in the field of endo- and boroscopic tests.

**Reference Architecture Model RAMI 4.0**

IIoT, OPC UA, DICONDE and the AAS are concepts for NDE 4.0. But how are they connected, which different tasks do they perform and how can they be located?

This task of locating Industry 4.0 concepts is fulfilled by the Reference Architecture Model for Industry 4.0 (RAMI 4.0) (DIN SPEC 91345:2016-04) (see Figure 11). Unfortunately, RAMI 4.0 is quite abstract, however it is one of the core models for Industry 4.0. Therefore, it is discussed shortly in the following.

RAMI 4.0 shows the Industry 4.0 world which has to be completely covered by interfaces. With the help of RAMI 4.0, every Industry 4.0 standard, interface, protocol, administration shell and every asset can be described and located in a structured way. RAMI 4.0 also helps to clarify whether all necessary interfaces exist.

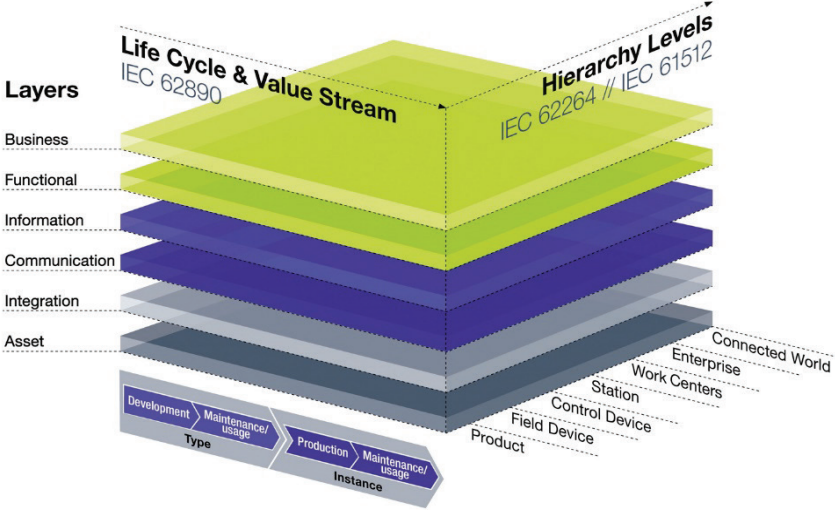

**Fig. 11** The Reference Architecture Model Industry 4.0 (RAMI 4.0)
©Platform Industry 4.0 (DIN SPEC 91345:2016-04)**.**

The Life Cycle & Value Stream axis represents the value chain and the life cycle of an asset, starting with the development and usage of a new type, through the production of the instance to the usage of the instance. The term "type" is used to identify a new asset type, such as a new X-ray inspection system. Instance refers to the test facilities that have actually been built.

The hierarchy levels correspond to the layers of the automation pyramid (refer to Figure 4) besides the top level "Connected World". The automation pyramid only covers communication within enterprises, however for Internet 4.0 data exchange between companies this layer needs to be included.

On the architecture axis (Layers) the lowest layer (Asset) represents the physical object. The "Integration Layer" is the transition layer between the physical and the information world. "Comunication", "Information" and "Functional Layer" are abstraction layers for the communication and the "Business Layer" describes the business perspective.

The Industrial Internet Reference Architecture (Industrial Internet Consortium 2019), published by the IIC, defines similar architecture layers compared to RAMI 4.0. However, it does not consider the other two axes.

*Location of AAS, IIoT, OPC UA and AutomationML to RAMI 4.0*

Due to the three axes design RAMI 4.0 is the ideal tool to locate all Industry 4.0 concepts.

OPC UA, like most communication protocols, covers the information and communication layers for instances (not for types), i.e. the right half of the middle two layers in Figure 9. Moreover, the connected world and the enterprise level is not covered by OPC UA.

Due to its connection gateways between different connectivity standards the IIoT Connectivity Framework covers the enterprise level, but not the connected world level.

AutomationML, an XML-based data format for storing and exchanging plant design data, covers the left half of the middle two layers in Figure 9. AutomationML therefore serves to describe the type of an asset.

The AAS sees itself as a virtual image, the digital twin, of each asset and thus as a link between all interfaces and protocols within the Industry 4.0 world. Projects for mapping between OPC UA, AutomationML and AAS have been started and will be detailed in future releases.

**Data Sovereignty, Data Markets, and Connected Internet 4.0 World**

As shown in Table 1, the networking of industrial production through standardized interfaces and thus the storage and use of the resulting crosslinked data sets is elementary for the fourth industrial revolution. However, the linked data records also represent a value in themselves. Data itself becomes an asset. There is a market for data and it is important to use it. The way to this market is NDE 4.0 with the interfaces discussed in this publication. How to make this market safe is and how to connect data between different companies is discussed in this section.

> *"The key focus for a data-driven economy*
> *and new business models is on linking data."*
> *[Quote: International Data Space Association]*

In the future, it will be possible to buy data independent of suppliers. The aim is to prevent illegal data markets, to create data markets according to crucial values (like data privacy and security, equal opportunities through a federated design, and ensuring data sovereignty for the creator of the data and trust among participants) and to ensure that companies that have generated the data also benefit from their value and not just a few large data platforms.

The International Data Space Association (IDSA) has set itself this goal. IDSA develops standards and de-jure-standards based on the requirements of IDSA members, works on the standardization of semantics for data exchange protocols and provides sample code to ensure easy implementation.

One of the key elements IDSA is implementing are the so-called IDS connectors (International Data Spaces Association 2019) which guarantee data sovereignty (see figure 12). Both the data source and the data sink have certified connectors. The data provider defines data use restrictions. The data consumer connector guarantees that the restrictions are followed. For example, if the data provider defines that the data consumer is allowed to view the data once the data will be deleted by the consumer connector after the data was viewed. This enables also the producer of the data to decide which customer can use his data in which form as an economic good, for statistical evaluation or similar.

Due to those connectors IDSA enables the connected world as required by RAMI 4.0.

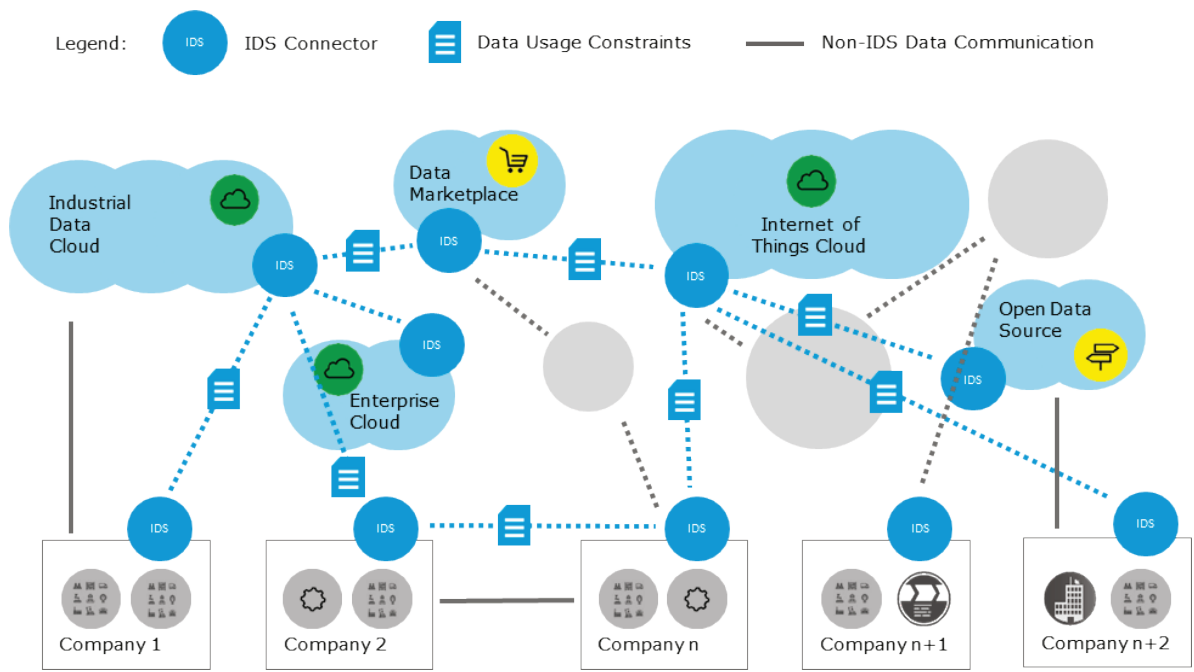

**Fig. 12** IDSA: Connected Industry 4.0 World (International Data Spaces Association 2019)**.**

For many, marketing the data will be a new business model. For NDE it is the opportunity to move from the position of an unnecessary cost factor to one of THE data suppliers. This will create a new, larger business case.

In order to help shape this development and equip NDE for the data market, DGZfP recently became a member of the IDSA.

**Summary and outlook**

With the AAS, IIoT, OPC UA, WebServices, AutomationML, and IDSA, protocols and interfaces have already been created in industry to implement NDE 4.0. In order to make NDE an integral part of the Industry 4.0 world, cooperation is required. Firstly, ontologies must be created for OPC UA (Companion Specifications), for WebServices (Web Ontology Language), for AutomationML and for the Asset Administration Shell to assure semantic interoperability. On the other hand, there is the task of guaranteeing the requirements of the NDE industry in the IDSA.

With DICOM/DICONDE there is also an advanced interface and a well-developed open data format available. DICOM/DICONDE already offers semantic interoperability and its standardized and open ontology can be used as a base for the NDE ontologies for the standard Industry 4.0 interfaces mentioned in the paragraph above.

For NDE technologies with large data volumes, DICONDE is an ideal addition to the industrial interfaces (similar to the combination HL7 and DICOM). This means that interfaces/mappings from DICONDE to the Industry 4.0 world (OPC UA) are needed. For NDE technologies with small data volumes, it is necessary to decide, depending on the application, whether a direct interface is created using OPC UA or whether these are first stored in the DICONDE world and then transferred to the OPC UA world, in order to summarize all test results in one place. In addition, it is necessary to check which steps are required to be able to use DICONDE for ultrasound and eddy current.

In general, a revision-safe and secure storage must always be ensured. The retrievability, integrity and sovereignty of the data is key. Most of those requirements are already implemented in DICONDE and OPC UA.

Other open data-formats for NDE data, like HDF5, can be seen as alternatives to DICONDE. However, for most inspection situations the standardized open information models of DICONDE, which enable machine readable data using semantic interoperability, surpass the information models of the other data formats. Also, revision-safe and secure data-storage needs to be implemented in addition.

In order to ensure the interests of NDE in the Industry 4.0 world and for the development of the necessary ontologies, cooperation with Industry 4.0 must be strengthened.

*NDE 4.0 is the chance for NDE to move from the niche of the "unnecessary cost factor" to one of the most valuable data providers for Industry 4.0. However, this requires the opening of data formats and interfaces. The insight that the protectionism lived up to now will have a damaging effect on business in the foreseeable future will decide on the future of individual companies. For companies that recognize the signs of the times, NDE 4.0 is also the way to the data market, to a completely new business model for the industry.*

**Acknowledgement**

Many thanks to Ripi Singh (Inspiring NEXT) and Daniel Kanzler (applied NDT Reliability) for all the discussions about NDE 4.0. Also to Jens Martin (VISUS Industry IT) for the introduction to DICONDE and HL7, to Thomas Usländer (Fraunhofer IOSB) for the information about the platform Industry 4.0, the AAS and OPC UA, to Markus Eberhorn (Fraunhofer EZRT) for the introduction to OPC UA and to Ralf Casperson (BAM) for a first study regarding the applicability of the DICONDE standard for eddy current applications.

Many thanks also to Sven Gondrom-Linke (Volume Graphics) for his work as Vice Chairman of the DGZfP subcommittee "Interfaces for NDE 4.0", to the members of the subcommittee, to the hosts of the meetings and to the participants of the survey on Facebook and LinkedIn. Last but not least I have to thank Franziska Vrana for the Industry and NDE 4.0 images and for all her support.